\pdfoutput=1

\documentclass[11pt]{article}
\usepackage{jheppub}

\usepackage{amsmath}
\usepackage{latexsym,amssymb}
\usepackage{graphicx,color,slashed}
\usepackage{bbm} 
\usepackage{mathtools}
\usepackage{ifpdf}

\newcommand{\cG}{\mathcal{G}}

\newcommand{\cE}{\mathcal{E}}

\newcommand{\cH}{\mathcal{H}}
\newcommand{\cL}{\mathcal{L}}

\newcommand{\cN}{\mathcal{N}}

\newcommand{\cV}{\mathcal{V}}

\newcommand{\be}{\begin{equation}}
\newcommand{\ee}{\end{equation}}
\newcommand{\ba}{\begin{eqnarray}}
\newcommand{\ea}{\end{eqnarray}}

\renewcommand{\a}{\alpha}

\def\E{{$E_{7(7)}$}}

\newcommand{\rf}[1]{(\ref{#1})}
\newcommand{\bea}{\begin{eqnarray}}
\newcommand{\eea}{\end{eqnarray}}

\def\bfzero{\relax{\rm I\kern-.18em 0}}
\def\bfone{\relax{\rm 1\kern-.35em 1}}
\def\twomat#1#2#3#4{\left(\begin{array}{cc}
\end{array}
\right)}

\def\o#1#2{{{#1}\over{#2}}}

\def\a{\alpha}

\def\ad{{\dot\alpha}}

\def\csN{{\fontsize{9.35pt}{9pt}\selectfont \mbox{$\cN$} \fontsize{12.35pt}{12pt}\selectfont }}
\def\cssN{{\fontsize{6.35pt}{6pt}\selectfont \mbox{$\cN$} \fontsize{12.35pt}{12pt}\selectfont }}

\def\o{\omega}\def\O{\Omega}

\usepackage[mathscr]{euscript}

\def\gh{\mathfrak{h}}
\def\gf{\mathfrak{f}}

\title{\rm{\bf    Superinvariants  Below Critical Loop Order  }}
\author
{ Renata Kallosh}
\affiliation{Stanford Institute for Theoretical Physics and Department of Physics, Stanford University, Stanford, CA 94305, USA}

\abstract{  Investigation of on-shell superinvariants is the standard way to identify the candidate UV divergences in supergravity.  Geometric on-shell superinvariants in 4d supergravity at $\cN\geq 4$  are available starting at the critical loop order $L=\cN$ \cite{Kallosh:1980fi,Howe:1980th}.  However,  more than 10 years 
 ago, it was conjectured that the UV divergencies in supergravity might appear earlier,  for $L=\cN-1$, due to the existence of harmonic superspace superinvariants based on  Grassmann analytic superfields \cite{Bossard:2011tq}. Here we show that these harmonic  superinvariants
 at the nonlinear level 
 are inconsistent, because Grassmann analyticity breaks local $\cH$ symmetry of the $\cN\geq 4$ on-shell superspace.  Therefore, the ``puzzling enhanced cancellation'' of UV divergences   for $L=\cN-1$  in superamplitude calculations for $ L=3, \, \cN=4$ and $L=4, \, \cN=5$  is explained by nonlinear
 local supersymmetry consistent with unbroken local $\cH$ symmetry.
}

\begin{document}

\maketitle

%\tableofcontents{}

%\newpage

\parskip 5pt

%%%%%%%%%%%%%%%%%%%%%%%%%%%%%%%%%%%%%%%%%%%%%%%%%%%%%%%%%%%%%%%%%%%%%%
%%%%%%%%%%%%%%%%%%%%%%%%%%%%%%%%%%%%%%%%%%%%%%%%%%%%%%%%%%%%%%%%%%%%%%
%%%%%%%%%%%%%%%%%%%%%%%%%%%%%%%%%%%%%%%%%%%%%%%%%%%%%%%%%%%%%%%%%%%%%%
%%%%%%%%%%%%%%%%%%%%%%%%%%%%%%%%%%%%%%%%%%%%%%%%%%%%%%%%%%%%%%%%%%%%%%
%%%%%%%%%%%%%%%%%%%%%%%%%%%%%%%%%%%%%%%%%%%%%%%%%%%%%%%%%%%%%%%%%%%%%%

\section{Introduction}

$\cN\geq 4$ supergravities with physical scalars in ${\cG\over \cH}$ coset space in 4d have a formulation with independent and linearly realized duality $\cG$ symmetry and local $\cH$-symmetry \cite{Cremmer:1979up, deWit:1982bul}. Local $\cH$-symmetry  is a $U(\cN)$ (or $SU(\cN)$) group. These symmetries are also present in the on-shell superspace for these theories \cite{Brink:1979nt, Howe:1981gz}.

In \cite{Kallosh:2023asd,Kallosh:2023thj,Kallosh:2023css} the assertion was made that for $\cN\geq 4$ supergravities\footnote{ $\cN\leq 4$  supergravities in 4d have anomalies but no anomalies are known at $\cN\geq 5$.} 
 nonlinear supersymmetry invariants cannot
be constructed unless 1) they come from a whole superspace integral, not a subspace of the whole superspace, which is necessary for the manifest local $\cH$ symmetry of the measure of integration  2) the geometric superspace Lagrangian respects both global $\cG$-duality symmetry and local $\cH$ symmetry. These conditions provide the definition of $L_{cr}=\cN$ for 4d supergravities at which nonlinear superinvariants can be constructed, based on dimensional analysis in on-shell superspace \cite{Kallosh:1980fi,Howe:1980th}. Therefore if there is a UV divergence at $L<L_{cr}$ it means that nonlinear supersymmetry is broken by this UV divergence.

Our goal here is to provide a detailed explanation of the statements made in \cite{Kallosh:2023asd,Kallosh:2023thj,Kallosh:2023css} with regard to $L=\cN-1$ superinvariants\footnote{It is known that for $\cN\geq 5$ UV divergences at $L\leq\cN-2$ would mean that E7 type $\cG$ symmetry is broken \cite{Beisert:2010jx}, \cite{Freedman:2018mrv}. However, the case of $L=\cN-1$ is not protected by the soft scalar limit theorem. For example, the case L=7 for $\cN=8$ or L=4 for $\cN=5$ are not protected by \E\, or $SU(1,5)$, respectively. But since at L=4,   $\cN=5$  UV divergences in 82 diagrams cancel  \cite{Bern:2014sna}, this requires an explanation why harmonic $L=\cN-1$ superinvariants predicting this UV divergence are not valid at the nonlinear level.}
 in an on-shell harmonic superspace proposed in \cite{Bossard:2011tq}. We argued in  \cite{Kallosh:2023asd,Kallosh:2023thj,Kallosh:2023css}  that these counterterms/superinvariants are valid at the linear level but break 
 nonlinear local supersymmetry.
 
 Our argument in \cite{Kallosh:2023asd,Kallosh:2023thj,Kallosh:2023css} was based on the fact that the proof of the consistency of the harmonic superspace $(\cN , 1, 1)$  proposed in \cite{Hartwell:1994rp} was given for Yang-Mills theory and for $\cN = 1, 2, 3, 4$ conformal supergravity, but not for   $\cN\geq 4$ Poincar\'e supergravity.  Here we will study the relevant  $L=\cN-1$ superinvariants of the harmonic superspace $(\cN , 1, 1)$ and show directly  that they are inconsistent at the nonlinear level. This will explain  why nonlinear supersymmetry invariants are not available at $L < L_{cr}$.

  An important issue to stress here is the closure of a local supersymmetry algebra:  for example, in the case of $\cN=8$ supergravity the classical action \cite{Cremmer:1979up, deWit:1982bul} has local supersymmetry, general covariance, local Lorentz symmetry, Maxwell symmetry and local $SU(8)$-symmetry (local 
$\cH$-symmetry). The commutator of two nonlinear local supersymmetries involves, in addition to all other local symmetries of the action, also a local $SU(8)$ symmetry, and a term proportional to equations of motion \cite{Cremmer:1979up, deWit:1982bul}
  \be
\{\delta_{ 1 s}, \delta _{ 2 s} \} = \delta _{\rm Diff} + \delta_{3 s}+ \delta _{SO_ {(3,1) }} +  \delta _{U(1) } +\delta _{SU(8)}  + {\rm EOM}
\label{susy}\ee
A detailed derivation of this commutator in the nonlinear case is presented in Sec. 3.4  in \cite{Hillmann:2009zf}. In the linear case, the term $\delta _{SU(8)}$ in the RHS is absent.
The reason we present here the local supersymmetry anti-commutator in \rf{susy}  is to show that breaking of local $SU(8)$ symmetry by quantum corrections means that nonlinear local supersymmetry is inevitably broken. This is also valid in case that equations of motion are satisfied, i.e. on-shell, when terms ${\rm EOM}$ in \rf{susy} vanish. Thus
\be
{\rm broken } \, \,  {\rm local } \, \, \cH \quad \Rightarrow \quad  {\rm broken }\, \,  {\rm nonlinear } \, \, {\rm supersymmetry}
\ee
In the unitary gauge, when local $SU(8)$ symmetry is gauge-fixed, the algebra of two supersymmetries still includes a field-dependent nonlinearly realized $SU(8)$ symmetry, with duality $\cG$ symmetry parameters.

We will also show in detail why the whole superspace integral measure of integration for $\cN\geq 4$ respects local $\cH$ symmetry. It is broken at the nonlinear level if any subspace of the whole superspace is used for superinvariants, unless new coordinates are added to a standard superspace.
If new coordinates, like harmonic ones, are used to define superinvariants, there is no problem at the linear level. The problem only appears at the nonlinear level at $\cN\geq 4$ as we will show.

The off-shell harmonic superspace construction for SYM theories and $\cN=2$ supergravity was described in  \cite{Galperin:2001seg}. We present a short description of these harmonic superspaces and projective superspaces \cite{Kuzenko:2008ep,Kuzenko:2008ry} in Appendix \ref{App:GIKOS}.

The important difference between $\cN=2$ and $\cN\geq 4$ is that at $\cN\geq 4$ there is 
a ${\cG\over \cH}$ coset structure and the relevant global $\cG$ and local $\cH$ symmetries are present.
In $\cN=2$ supergravity superspace, originating from conformal $\cN=2$ superspace,   it is possible to gauge-fix local tangent space $\cH=U(2)$ symmetry of the standard superspace, see  \cite{Howe:1981gz}, sec. 7 on ``de-gauging'' local $U(2)$ symmetry. This local symmetry was present in  $\cN=2$ conformal supergravity but was eliminated in Poincar\'e $\cN=2$ supergravity. However, at higher $\cN$, the local $(U(\cN)$ $\cH$ symmetry is a tangent space symmetry of the superspace  \cite{Howe:1981gz}; it is at the core of the superspace construction.

Despite this fact, it was assumed in \cite{Bossard:2011tq} that to define on-shell harmonic superspace, one only needs to satisfy integrability conditions for Grassmann analyticity conditions. It was indeed shown in \cite{Bossard:2011tq}  that these are satisfied for $L=\cN-1$ type supersymmetry invariants.  

We will show here that these integrability conditions are necessary but not sufficient for the consistency of the $\cN \geq 4 $ harmonic on-shell superspace. They are actually sufficient in $\cN=2$ case, see \cite{Kuzenko:2008ep,Kuzenko:2008ry} whose strategy was used in \cite{Bossard:2011tq} to construct on-shell harmonic superspace at $\cN \geq  4$.

But in  $\cN\geq 4$ there is a duality $\cG$-symmetry which enforces a compensating field dependent $(S)U(\cN)$ transformation, in case that local $\cH$-symmetry is gauge-fixed in the spacetime action \cite{Cremmer:1979up, deWit:1982bul,Kallosh:2008ic,Bossard:2010dq}. This compensating field-dependent $(S)U(\cN)$ transformation
 is incompatible with harmonic coordinates. Therefore, the Grassmann analyticity in on-shell harmonic superspace assumed in \cite{Bossard:2011tq} for $\cN \geq 4$ supergravities breaks local $(S)U(\cN)$ symmetry and nonlinear supersymmetry.

In $d>4$ the relevant ${\cG\over \cH}$ supergravity superinvariants below critical loop order are also known in harmonic superspace at the linearized level \cite{Bossard:2009sy,Bossard:2014lra,Bossard:2014aea,Bossard:2015uga}. These are useful to describe the UV divergences which did occur at $d>4$. None of these harmonic superinvariants with Grassmann-analytic superfields were claimed to exist at the nonlinear level.  In all cases, they are at $L<L_{cr}$ \cite{Kallosh:2023css} and show that local $\cH$ symmetry and nonlinear supersymmetry are broken by quantum corrections. This means that these local symmetries are anomalous. We explain in Appendix \ref{App:C} how the arguments in case of 4d supergravities with $ L=\cN-1$  work for $d>4$ supergravities and obstruct the nonlinear generalization of the harmonic superinvariants constructed in \cite{Bossard:2009sy,Bossard:2014lra,Bossard:2014aea,Bossard:2015uga}.

\section{Local $\cH$ symmetry of the whole curved superspace volume}

The value of  $L_{cr}$  in integer dimensions $d\geq 4$ was determined  by dimensional analysis in  superspace, in  4d it is $L_{cr}= \cN$ \cite{Kallosh:1980fi,Howe:1980th}. It was based on the on-shell superspace construction in  \cite{Brink:1979nt, Howe:1981gz}. The ones below $L_{cr}$ are either defined in a subspace of the whole superspace, which breaks local $\cH$ symmetry or depend on Lagrangians breaking $\cG$ symmetry. In $d>4$ the value of $L_{cr}$ was determined in \cite{Kallosh:2023css}.

The whole superspace has the measure of integration, which has both curved space symmetry as well as tangent space symmetry.
\be
dV =  d^d x \, d^{4\cN} \theta \, {\rm Ber} \, E  \ .
\label{vol}\ee
Here, Berezinian, denoted ${\rm Ber}\, E$,   is a generalization of the determinant of the curved space in the presence of Grassmann coordinates, where it is defined as a superdeteminant.
This is analogous to  general relativity where $dx^m e_m^a(x)= E^a$, and the volume of integration can be given in the form with manifest curved space symmetry or in a form with manifest tangent space Lorentz symmetry $SO(1.3)$
\be
\int d^4 x \sqrt g \quad \Rightarrow \quad \int E^a\wedge E^b \wedge E^c\wedge E^d \, \epsilon_{abcd}  \ .
\ee
Note that the Lorentz space covariant tensor $\epsilon_{abcd}$ needs four of $E^a$ to be saturated.
An example of a subspace of a  whole space breaking tangent space symmetry in pure gravity in 4d  would be an integral over less than four $E^a$'s. In this case, the local $SO(1.3)$ Lorentz symmetry of the measure of integration would be broken.

The  4d curved superspace coordinates  $Z^M= (x^m, \theta^{\underline \mu}, \theta^{\underline {\dot \mu}} ) $ include in addition to spacetime coordinates $x^m$ also $4\cN$ anticommuting coordinates $\theta^{\underline \mu}, \theta^{\underline {\dot \mu}}$. The relevant preferred frames are defined as 
\be
E^A= dZ^M E_M{}^A (Z)
\ee
There are general coordinate transformations  $Z^M \rightarrow Z^M + \xi^M(Z)$ and tangent space rotations: 
\be 
\delta E^A= E^B L_B{}^A
\label{tang}\ee 
These tangent space symmetries play an important role in the construction of superinvariants. Here $A= (a, {\underline \alpha}, {\underline {\dot \alpha}})$ where 
\be
E^{\underline \alpha} = E^\alpha_i \qquad  E^{\underline {\dot \alpha} } = E^{ \dot \alpha \, i}\, ,  \qquad \a, \dot \a =1,2 \, , \qquad i=1, \dots, \cN
\ee
The tangent space group (also called structure group) is a product of a local Lorentz group and local internal group $SL(2, \mathbb{C}) \times \cH$ where $\cH$ is $U(\cN)$ or its subgroup. This means that $SL(2, \mathbb{C}) \times \cH$ parameters in \rf{tang} are functions of the superspace coordinates $Z^M= (x^m, \theta^{\underline \mu}, \theta^{\underline {\dot \mu}} ) $.
\be 
 L_B{}^A= L_B{}^A(Z) \ .
\ee
Let us also define local $\cH$ tensors
\be
E_{ij} = E^\alpha_i \wedge E^\beta_j \epsilon_{\alpha \beta}\, ,  \qquad \bar E^{kl} = \bar E_{\dot \alpha}^k  \wedge \bar E_{\dot \beta}^l\epsilon^{\dot \alpha \dot \beta} 
\ee
and their products
\be
E_{i_1\dots i_{\cN} j_1\dots j_{\cN}}  = E_{i_1j_1} \wedge \dots \wedge E_{i_{\cN} j_{\cN}} \ ,
\ee
\be
\bar E^{i_1\dots i_{\cN} j_1\dots j_{\cN}}  = \bar E^{i_1j_1} \wedge \dots \wedge \bar E^{i_{\cN} j_{\cN}} \ .
\ee
The integral of the whole superspace volume \rf{vol} can be presented in the form
\bea
 \int E^a\wedge E^b \wedge E^c\wedge E^d \, \epsilon_{abcd} \, \, 
E_{i_1\dots i_{\cN} j_1\dots j_{\cN}} \epsilon^ {i_1\dots i_{\cN}} \, \epsilon^ {j_1\dots j_{\cN}}\, \, 
 \bar E^{l_1\dots l_{\cN} k_1\dots k_{\cN}} \epsilon_ {l_1\dots l_{\cN}} \, \epsilon_ {k_1\dots k_{\cN}} \ ,
\label{vol1}\eea
where $SL(2, \mathbb{C}) \times \cH$ symmetries with
parameters which are functions of the superspace coordinates $Z^M= (x^m, \theta^{\underline \mu}, \theta^{\underline {\dot \mu}} ) $ are manifest.

Clearly, any attempt to integrate over less of $E^\alpha_i$, $\bar E_{\dot \alpha}^k$ than $4\cN$ will break $U(\cN)$ $\cH$-symmetry: to saturate $\epsilon^ {i_1\dots i_{\cN}} \, \epsilon^ {j_1\dots j_{\cN}} \epsilon_ {l_1\dots l_{\cN}} \, \epsilon_ {k_1\dots k_{\cN}}$ one needs $4\cN$ of these. This is the same as in 4d: to preserve  local  Lorentz symmetry, one needs 4 of $E^a$'s to 
saturate $ \epsilon_{abcd} $.

As long as no new coordinates are introduced into superspace, the only $SL(2, \mathbb{C}) \times \cH$ invariant measure of integration has dimension $-d+2\cN$, in 4d case
\be
{\rm dim} [dV_{whole}]= -4+2\cN \ .
\ee 
But when harmonic coordinates are added to a superspace, at the linear level, this opens a possibility to have the dimension of the measure of integration less than $-d+2\cN$. 
Specifically, in the case of \cite{Bossard:2011tq} and in 4d $ \cN\geq 4$, the Grassmann analytic subspace of the superspace has a measure of dimension
\be
dim [dV_{harm}]= -4+2(\cN-1) \ ,
\ee 
which explains the claimed existence of $L=\cN-1$ harmonic superinvariants. We will see that at the linear level, these are consistent; however, we will show that the local $\cH$-symmetry of the measure of integration over the subspace is broken at the nonlinear level.

\section{ $\cN\geq 4$ on-shell superspace geometry and its tangent space $\cH$ symmetry}\label{Sec:3}

The physical scalars in these theories are in the ${\cG\over \cH}$ coset space where
where $\cH$ is $U(\cN)$ or its subgroup, e. g. $SU(8)$ for $\cN=8$, $U(6)$  for $\cN=6$, $U(5)$  for $\cN=5$, $U(4)$  for $\cN=4$. The groups $\cG$ are \E\,   for $\cN=8$, $SO^*(12)$ for $\cN=6$ and $SU(5,1)$ for $\cN=5$, $SU(4)\times SU(1, 1)$ for $\cN = 4$.  These are known as groups of type E7.

Following \cite{Howe:1981gz} we consider a superspace with coordinates 
\be
z^{\mathscr{M}}= ( z^{M},  y^{\underline{ m}}, y^{\underline {m'}} ) \ee 
where $z^M$ are the
coordinates of ordinary superspace with four bosonic coordinates $x^m$ and $2\cN$ anticommuting coordinates $\theta^{\underline \mu}$  and its complex conjugate $\theta^{\underline {\dot \mu}}$. This is a standard superspace where $\cN=8$ on-shell supergravity was studied in \cite{Brink:1979nt}. 
In  \cite{Howe:1981gz}, the extra bosonic coordinates were added:
 $y^{\underline{ m}}$  are ${1\over 2} \cN(\cN- 1)$ complex coordinates with complex conjugates $y^{\underline {m'}}$. This version is given (mostly in Sec. 8) of \cite{Howe:1981gz} for all $\cN$, and its geometry is consistent with a normal superspace version with coordinates 
$z^{M}$ in \cite{Brink:1979nt} where $\cN=8$ was studied.

A tangent space vector is
\be
X^{\mathscr{A}}= ( X^{A},  X^{\underline{ a}}, X^{\underline {a'}} ) \qquad X^{A} = (X^a, X^{\underline{ \a}}, X^{\underline {\a'}})
\ee
\be
X^{\underline{ a}}= X_{ij}, \qquad X^{\underline {a'}}= \overline {X}^{ij}
\ee
The tangent space group is $SL(2, C) \times \cH$. Here $\cH$ is $SU(8)$ for $\cN=8$, $U(\cN)$  for $\cN=6,5,4$.
The matrix $L_{{\mathscr{A}}}{}^{\mathscr{B}}$ defining the tangent space rotation on the tangent vector  
\be
\delta X^{\mathscr{A}} = X^{\mathscr{B}} L_{{\mathscr{B}}}{}^{\mathscr{A}}(z^{\mathscr{M}})
\label{Lor_H}\ee
is given in eqs. (2.5),  (8.4) in \cite{Howe:1981gz}. 
It is convenient to use this version where all geometric fields depend trivially on extra coordinates
\be
D_{\underline{ a}} = \partial _{\underline{ m}}=0 \ .
\label{der} \ee
The vielbein one-form is
$
E^{\mathscr{A}} = dz^{\mathscr{M}} E_{\mathscr{M}}{}^{\mathscr{A}} 
$
and ${\mathscr{E}}^A= E^A$ due to \rf{der}.
There is a standard general coordinate transformation in a superspace. The superspace has a local tangent space symmetry $SL(2, C) \times \cH$ where $\cH$ is  $U(\cN)$ or its subgroup.
\be
\delta E^{\mathscr{A}} = E^{\mathscr{B}} L_{{\mathscr{A}}}{}^{\mathscr{B}}(z^{\mathscr{M}})
\label{tangent}\ee
Also, a one-form connection $\hat \Omega$ is introduced, which transforms inhomogeneously 
\be
\delta \hat \Omega _{{\mathscr{A}}}{}^{\mathscr{B}} = - d L_{{\mathscr{A}}}{}^{\mathscr{B}}- L_{{\mathscr{A}}}{}^{\mathscr{C}} \hat \Omega _{{\mathscr{C}}}{}^{\mathscr{B}} + \hat \Omega _{{\mathscr{A}}}{}^{\mathscr{C}} 
 L_{{\mathscr{C}}}{}^{\mathscr{B}} = - D L_{{\mathscr{A}}}{}^{\mathscr{B}} \ .
 \ee
The torsion two-form is an exterior derivative of  $E^{\mathscr{A}}$, and $R_{{\mathscr{A}}}{}^{\mathscr{B}}$ is the curvature two-form
\be
T^{\mathscr{A}}= D E^{\mathscr{A}}\,  , \qquad R_{{\mathscr{A}}}{}^{\mathscr{B}}= d \hat \Omega _{{\mathscr{A}}}{}^{\mathscr{B}}  + \hat \Omega _{{\mathscr{A}}}{}^{\mathscr{C}} \wedge \hat \Omega _{{\mathscr{C}}}{}^{\mathscr{B}} \ .
\ee
This leads to Bianchi identities 
\be
DT^{\mathscr{A}}= E^{\mathscr{B}} \wedge R_{{\mathscr{A}}}{}^{\mathscr{B}}\, , \qquad  D R_{{\mathscr{A}}}{}^{\mathscr{B}}=0 \ .
\label{BI}\ee
The set of fundamental Bianchi identities \rf{BI} was solved in \cite{Brink:1979nt,Howe:1981gz} where the set of on-shell constraints was imposed on solutions.

An important point here is the role played by the global E7-type  symmetry for $\cN\geq 4$. The tangent space symmetries in eq. \rf{Lor_H} include  local Lorentz symmetry and local $(S)U(\cN)$ $\cH$-symmetry. The superspace connection $\hat \Omega$ is constructed from the superspace vielbein
\begin{eqnarray}\label{vb}
{\cal V} =\left(
                                        \begin{array}{cc}
                                          U^{IJ}{} _{ij}  & \bar V^{IJij} \\
                                           V_{IJij} &  \bar U_{IJ}{}^{ij} \\
                                        \end{array}
                                      \right)\ , 
                                      \end{eqnarray}
where ${\cal V}$ an element of the Lie group $\cG$: $(S)U(\cN)$ acts on the small indices $ij$ in a natural way with superspace-depending parameters, while the global group $\cG$  acts on the capital indices by
\be
\delta {\cal V} = -\Lambda {\cal V}\, ,  \qquad 
\Lambda = \left(
                                        \begin{array}{cc}
                                          \Lambda^{IJ} {} _{KL}& \, \,  \bar \Sigma ^{IJKL} \\
                                          \Sigma _{IJKL} & \, \,  \bar \Lambda _{IJ}{}^{KL} \\
                                        \end{array}
                                      \right)\ , 
                                   \label{Lambda}\ee
The superspace connection $\hat \Omega=(d {\cal V}^{-1}) {\cal V}$ is  $\cG$-invariant                                                      
and it is a connection one-form having zero curvature                                      
$
d\hat \Omega + \hat \Omega \wedge \hat \Omega=0
$.
This summarizes the status of the $\cN\geq 4$ on-shell superspace \cite{Brink:1979nt,Howe:1981gz}.

Is is possible to gauge-fix the tangent space $U(\cN)$ symmetry of the superspace with $\cN\geq 4$? We show how this is done by ``de-gauging'' in $\cN=2$ supergravity in Appendix \ref{App:GIKOS} where there is no relevant duality $\cG$-symmetry.

Such a ``de-gauging'' was not done in $\cN\geq 4$ superspace, but the gauge-fixing of $\cH$ local symmetry was performed in the unitary gauge for the spacetime Lagrangian in \cite{Cremmer:1979up, deWit:1982bul,Kallosh:2008ic,Bossard:2010dq}. It explains why the on-shell superspace \cite{Brink:1979nt,Howe:1981gz} needs to preserve the  local $\cH$  symmetry as a tangent space symmetry.

\section{Harmonic on-shell superspace of $\cN\geq 4$ supergravities }
The on-shell harmonic superspace counterterms for $\cN\geq 4$ supergravities were constructed in \cite{Bossard:2011tq} and have a universal form for $\cN\geq 4$.
The superinvariants in the harmonic superspace at $d=4, L=\cN-1$ are one loop order below $L_{cr}=\cN$ and represent a ${1\over \cN}$ BPS invariants. This means that instead of the whole superspace integrals $d^{4\cN }\theta$, they involve
$d^{4(\cN-1)}\theta$ integration.

In 4d $\cN$-extended on-shell superspace was defined in  \cite{Brink:1979nt, Howe:1981gz}, and we described it shortly in Sec. \ref{Sec:3}.
The purpose of introducing additional coordinates, the harmonic ones, is to be able to justify subspace integrals over a fraction of the whole superspace and to define the so-called Grassmann analytic superfields, which depend on the coordinates of the analytic subspace only.

The off-shell harmonic superspace is not known to exist for supergravity with $\cN> 2$ \cite{Galperin:2001seg}, but for the case of interest with $\cN\geq 4$, the hope was that the on-shell harmonic superspace could be constructed. So the proposal for the on-shell harmonic superspace of supergravity was to add harmonic coordinates $u^I{}_i$ and their inverse $u^i{}_I$, $i=1, \dots, \cN, \, I=1, \dots, \cN$. The role of these coordinates is to represent the flag manifold $H\backslash K$
\be
\mathbb{F}_{1,1} (\cN) \cong \bigl( U(1)\times  U(\csN-2)\times U(1)\bigr)\backslash U(\csN)  \ ,
\ee
where  $K\cong U (\cN )$ and the 
isotropy group is $H\cong U(1)\times  U(\csN-2)\times U(1)$. It is proposed in \cite{Bossard:2011tq}, with reference to harmonic spaces in \cite{Galperin:2001seg}, to work in $\cN\geq 4$ supergravities on such a coset space. The  index $I$ is split according to the structure of the isotropy group: 
\be
I=(1,r,\csN)
\ee
The element of $U(\csN)$ is denoted by $u^I{}_i$, and it is stated that {\it the local gauge group acts to the right and the isotropy group acts to the left}. 
 At this point, we need to clarify the meaning of the element of $U(\csN)$ called $u^I{}_i$ and defined in a proposal made in \cite{Bossard:2011tq}. 
 
In the principal  $U(\cN)$ bundle  there is a Lie-algebra-valued one-form $\omega$  that combines the
Maurer-Cartan form on the group with the $U(\cN)$ connection on the base: 
\be
\omega= du u^{-1} + u\Omega u^{-1} \qquad \Rightarrow  \qquad \omega^I{}_J = du^I{}_i (u^{-1})^i{}_J + u^I{}_i \Omega^i{}_j  (u^{-1})^j{}_J \ .
\ee
Here $\O^i{}_j $ is the $U(\cN)$ connection of the standard superspace \cite{Brink:1979nt, Howe:1981gz}.
This form is  split  into its isotropy $\gh$ and coset components $\gf$
\be
\omega= \omega_{\gh}+ \omega_{\gf} \ .
\ee
The curvature is introduced
\be
d\o+\o^2= uRu^{-1} \ ,
\ee
and it is split into its $\gh$ and $\gf$ part:
\bea
D\o_\gf&=&-(\o_\gf\wedge\o_\gf)_\gf+(uRu^{-1})_\gf\nonumber\\
d\o_\gh+\o_\gh^2&=&-(\o_\gf\wedge\o_\gf)_\gh +(uRu^{-1})_\gh\ ,
\label{split}\eea
 $D$ here is the exterior derivative covariant with respect to $H$. It is suggested in \cite{Bossard:2011tq} that ``in  equations \rf{split}, the gauge was fixed  with respect to the isotropy group acting on $K$ so that $u$ are functions of local coordinates, $t$ say, on $F$.'' However, there is also a statement in \cite{Bossard:2011tq}: ``there has not been a choice of $U(\cN)$ gauge''.  We will study below  the consequences of the gauge-fixing performed in  \cite{Bossard:2011tq}.

The standard superspace \cite{Brink:1979nt, Howe:1981gz} has $\cN^2$ local symmetries of  $\cH=U(\cN)$ symmetry in the tangent space. The local tangent symmetry $SL(2, \mathbb{C}) \times  U(\cN)$ includes the local Lorentz symmetry as well as a local $U(\cN)$.  

The claim in \cite{Bossard:2011tq} is that the tangent space local symmetry of the harmonic space is $SL(2, \mathbb{C}) \times  H$, which means that the  local symmetries in the coset directions were gauge-fixed. Consider this in more details.

 The $\cH$ symmetries  are split in a harmonic superspace \cite{Bossard:2011tq} where harmonic coordinates are in a coset space 
\be
{\cH\over H}= {U(\csN)\over U(1)\times  U(\csN-2)\times U(1)}
\label{coset}\ee
There are $(\cN-2 )^2+2$  symmetries in $H$ and remaining 
\be
\cN^2-[(\cN-2 )^2+2]= 4\cN-6
\ee
in the coset direction. Thus, in presence of harmonic coordinates the local $\cH$ symmetry of the normal on-shell superspace is represented by $(\cN-2 )^2+2$ symmetries in $H$ and $4\cN-6$ symmetries in the coset space \rf{coset}.

The number of directions in the coset space \rf{coset} is $4\cN-6$. The harmonic measure of integration given in \cite{Bossard:2011tq} is
\be 
  \label{e:mu811}
  d\mu_{\scriptscriptstyle    (\cssN,1,1)}:=d^4x\, d^{4\cssN-6} t \, d^{2(\cssN-1)}\theta\, d^{2(\cssN-1)}\bar\theta\,\cE(\hat x)V(t)\ .
\ee
where additional $(4\cN-6)$ bosonic $t$-coordinates are present in the measure of integration 
and  $\hat x$ stands for all the harmonic superspace coordinates aside from 4 fermionic directions missing in \rf{e:mu811} comparative to whole $d^{4\cN} \theta$.

It was suggested in \cite{Bossard:2011tq} to introduce a  quantity $h(I)$  with $I=(1,r,\csN)$ such that 
\be h(1) = 1\ , \quad h(r) = 0 \ , \quad h(\csN) = -1 \ . \ee
The coset indices are then pairs $I,J$ such that $h(I)\neq h(J)$  and the coset  symmetry parameters are 
\be
(\Lambda ^I{}_J)_{\gf}\, : \qquad h(I)\neq h(J) \ .
\ee
while the $H$-indices are pairs $I,J$ with $h(I)=h(J)$ and the $H$ symmetry parameters are
\be
(\Lambda ^I{}_J)_{\gh}\, : \qquad h(I)= h(J) \ .
\ee
In standard on-shell superspace  \cite{Brink:1979nt, Howe:1981gz} with $\cN\geq 4$ all geometric torsion and curvatures depend on spinorial superfield $ \chi_{\alpha\, k ij} (x, \theta, \bar \theta)$, its conjugate,  and their superspace derivatives.

Now we  are looking at curvature constraints in harmonic superspace \cite{Bossard:2011tq}.  For example, in $\cN=4,5,8$, certain components of harmonic superspace curvature are quadratic in spinorial superfield $ \chi_{\alpha\, k ij} (x, \theta, \bar \theta)$ of the standard superspace and quadratic in harmonic coordinates, and take the following form
\be
(R^1_{ \a \, \dot \beta \cN})^1{}_1= (R^1_{ \a \, \dot \beta \cN})^{\cN}{}_{\cN} = {1\over 2} B_{\a \, \dot \beta} \neq 0 \ .
\label{Hcurv}\ee
Here
\be
B_{\a \, \dot \beta}(x, \theta, \bar \theta , u)= u^1{}_l \, \Big (\bar\chi_{\dot\beta}^{lij}(x, \theta, \bar \theta) \chi_{\alpha\, k ij} (x, \theta, \bar \theta) \Big) \, u^k{}_{\cN} \ .
\label{B}\ee

There is a subset of vector fields
$(\tilde E_\a^1,\tilde E_{\ad \cN}):=\tilde E_{\hat \a} $ and the corresponding curvature components of the harmonic superspace consists of $H$ and coset directions 
\be
(R_{\hat \a \, \hat \beta})^I{}_J= (R_{\hat \a \, \hat \beta})^I{}_J|_{\gh} + (R_{\hat \a \, \hat \beta})^I{}_J|_{\gf}
\ee
where $(R_{\hat \a \, \hat \beta})^I{}_J|_{\gh}$ has $h(I)= h(J)$ and 
$(R_{\hat \a \, \hat \beta})^I{}_J|_{\gf}$ has $h(I)\neq h(J)$. The constraints on these curvature components are given in \cite{Bossard:2011tq} in the form
\be
(R_{\hat \a \, \hat \beta})^I{}_J|_{\gf}=0\, 
\label{vanish}\ee
and 
\be
 (R_{\hat \a \, \hat \beta})^I{}_J|_{\gh}\neq 0 \qquad \Rightarrow \qquad (R^1_{ \a \, \dot \beta \cN})^1{}_1= (R^1_{ \a \, \dot \beta \cN})^{\cN}{}_{\cN} = {1\over 2} B_{\a \, \dot \beta} 
\label{Nvanish}\ee
based on the standard on-shell superspace structure in \cite{Howe:1981gz} 

 We see, therefore, from \rf{Hcurv}, \rf{B}  that  components of the on-shell harmonic superspace curvature, which are not vanishing, are in the  direction $(R_{\hat \a \, \hat \beta})^I{}_J|_{\gh}$ with $h(I)= h(J)$. All other components in direction $(R_{\hat \a \, \hat \beta})^I{}_J|_{\gf}$ with $h(I)\neq h(J)$ have to vanish, according to \rf{vanish}.

A tangent space symmetry for a curvature 2-form $R^I{}_J$ is
\be
 (R^I{}_J )'= \Lambda^I{}_K R^K{}_L \Lambda^L{}_J
\ee
This splits into the coset part and the $H$ part.
The coset part has the following symmetries 
\be
\Lambda^1{}_r,\, \, \Lambda^r{}_\cN,\, \, \Lambda^1{}_\cN
\ee
and the  complex conjugate ones are
\be
\Lambda^r{}_1,\, \, \Lambda^\cN{}_r,\, \, \Lambda^\cN{}_1
\ee
 Consider 3 cases  
\be
(R^1{}_{r} )'= \Lambda^1{}_K R^K{}_L \Lambda^L{}_{r} =  \Lambda^1{}_1 R^1{}_1 \Lambda^1{}_{r} +  \Lambda^1{}_{\cN} R^{\cN} {}_{\cN}  \Lambda^{\cN} {}_{r}+\dots
\, , 
\ee

\be
\, \, \, (R^r{}_{\cN} )'= \Lambda^r{}_K R^K{}_L \Lambda^L{}_{\cN}= \Lambda^r{}_1 R^1{}_1 \Lambda^1{}_{\cN} +\Lambda^r{}_{\cN} R^{\cN}{}_{\cN}\Lambda^{\cN}{}_{\cN} +\dots\, ,  
\ee

\be
\, \, \,  (R^1{}_{\cN} )'= \Lambda^1{}_K R^K{}_L \Lambda^L{}_{\cN}= \Lambda^1{}_1 R^1{}_1 \Lambda^1{}_{\cN} +\Lambda^1{}_{\cN} R^{\cN}{}_{\cN}\Lambda^{\cN}{}_{\cN} +\dots\, . 
\ee
All terms with $\dots$ vanish since they include curvatures vanishing according to \rf{vanish}. Since each of the vanishing coset curvature components is transformed into a non-vanishing $H$ curvature components $R^1{}_1=R^{\cN}{}_{\cN}\neq 0$, we find $2\cN-3$ complex conditions on parameters $\Lambda^I_{J}$, which is a condition  that $\Lambda^I_{J}|_{\gf}=0$. It means that there are $4\cN-6$ constraints on  local $U(\csN)$ symmetry of the tangent space
$\Lambda^i{}_j$
\be
\Lambda^I_{J}|_{\gf} = u^I{}_i \Lambda^i{}_j u^j{}_J=0
\label{Con}\ee
To summarize, the standard superspace \cite{Brink:1979nt, Howe:1981gz} has a local $U(\csN)$ symmetry of the tangent space with $\cN^2$ independent parameters $\Lambda^i{}_j(x, \theta, \bar \theta)$. But in the harmonic space with $R^I{}_J|_{\gf}=0$ and $R^1{}_1=R^{\cN}{}_{\cN}\neq 0$, there is a constraint \rf{Con} which says that all $4\cN-6$ coset symmetries are broken.

Curvatures in coset direction $R^I{}_J|_{\gf}=0$ have to vanish to provide an integrability condition for Grassmann analyticity. The only way to make this picture consistent is to break the coset symmetries associated with the coset space in \rf{coset}, i. e. to break  local $U(\csN)$ symmetry of the standard superspace of $\cN\geq 4$  supergravities in 4d.

Note that in $\cN\geq 4$ case there is no superspace with partially gauge-fixed $U(\cN)$ symmetry. In the next section, we explain the reason for this in the context of the ${\cG\over \cH}$ coset space of $\cN\geq 4$ supergravities \footnote{
In $\cN=2$ supergravity, this procedure of constructing harmonic superspace is consistent since in standard $\cN=2$ superspace, the local $U(2)$ symmetry can be gauge-fixed, using the compensating multiplets, as it was done in \cite{Howe:1981gz} or in projective superspace in \cite{Kuzenko:2008ep,Kuzenko:2008ry}. We stress here that in $\cN=2$  there is no spinorial superfield $ \chi_{\alpha\, [k ij]} (x, \theta, \bar \theta)$ from which non-vanishing curvature components are build as shown in eq. \rf{Nvanish}, and therefore all components of curvature $R^I{}_J$ vanish in $\cN=2$: local $U(2)$ is not a symmetry of the tangent space.}.

\section{Unitary gauge in spacetime with gauge-fixed local $\cH$-symmetry}\label{gf}
In spacetime, local $\cH$-symmetry was gauge-fixed for the classical Lagrangian in the original papers \cite{Cremmer:1979up, deWit:1982bul} where it was explained that a field-dependent compensating $SU(8)$ transformation is required to preserve the gauge.
A detailed study of this compensating $SU(8)$ transformation was performed in \cite{Kallosh:2008ic,Bossard:2010dq}.  
Under the combined action of local $SU(8)$ and global \E\,  the 56-bein transforms as
\be
{\cal V} \quad \rightarrow \quad {\cal V}' = h(x) {\cal V} g^{-1}\, ,  \qquad \qquad 
h(x)\in SU(8) \;, \;\; g\in E_{7(7)} 
\ee
i. e. from the left by a local $SU(8)$ and from the right by a global $E_{7(7)} $. Before gauge-fixing local 
$SU(8)$ the number of scalars is 133, the number of the elements in ${\cal V}$.

The unitary gauge is defined as follows \cite{Cremmer:1979up, deWit:1982bul}
\be
{\cal V} = {\cal V}^{\dagger}
\label{gauge}\ee 
which leaves us with 70 scalars $\phi^{ijkl}$ in the ${\bf 70}$ of $SU(8)$ \be
{\cal V}(x) \equiv  \exp \Phi(x)  =
\exp\left( \begin{array}{cc}  0  &  \phi_{ijkl}(x) \vspace{0mm}  \\
  \phi^{ijkl}(x)  & 0 \end{array} \right) 
\label{gauge1}\ee
with  
$\phi^{ijkl} = (\phi_{ijkl})^*$. The 133-70=63 local parameters of $SU(8)$ were used to eliminate the unphysical 63 scalars. 
This gauge is known as a unitary gauge, with only 70 physical scalars. 
The unitary gauge ${\cal V}= {\cal V}^{\dagger}$ is not preserved by a $\cG$-symmetry transformation since this symmetry involves all 133 scalars, elements of unconstrained ${\cal V}$. However, if a $\cG$-symmetry transformation 
 is complemented by a field-dependent $\cH$-symmetry transformation, it is possible to maintain the gauge. In such a case,
in the unitary gauge, the $\cH$-symmetry transformation is nonlinearly realized and depends on $x$-dependent scalar fields of the theory and on global parameters of $\cG$-symmetry.

The analysis in  \cite{Kallosh:2008ic} was performed in terms of  inhomogeneous coordinates of the ${E_{7(7)}\over SU(8)}$ coset space  defined as function of the independent scalars fields $\phi, \bar \phi$ as follows:
\be\label{yphi}
y_{ij, kl}(x)\equiv  \phi_{ijmn}(x)\left({\tanh(\sqrt {  \bar\phi \phi(x)}\over \sqrt {\bar  \phi \phi(x)}} \right)^{mn}_{\; \; \; \;\; {kl}}
\ee
Infinitesimal $E_{7(7)}$ transformation is
\begin{eqnarray}\label{Einverse}
E^{-1}=\left(
         \begin{array}{cc}
           1+\Lambda & -\Sigma \\
           -\bar{\Sigma} & 1+\bar{\Lambda} \\
         \end{array}
       \right)\ ,
\end{eqnarray}
In terms of the nonlinear scalar fields $y, \bar y$,  the matrix $\cV$ is given by the following expression
\begin{eqnarray}\label{Vy}
\cV(y, \bar y)=\left(
               \begin{array}{cc}
                 P^{-1/2} & -P^{-1/2}y \\
                 -\bar{P}^{-1/2}\bar{y} & \bar{P}^{-1/2} \\
               \end{array}
             \right)\, , \qquad P(y,\bar{y})_{ij}{}^{kl}\equiv(\delta_{ij}{}^{kl} -y_{ijrs}\bar{y}^{rskl})
\end{eqnarray}
The unitary  gauge condition (\ref{gauge}), \rf{gauge1} is preserved if the transformation law includes {\it simultaneous } multiplication of the 56-bein from the right by an arbitrary rigid $E_{7(7)}$ group element depending on 63 rigid parameters  $\Lambda_{J}{}^{L}$ and 70 rigid parameters $\Sigma_{IJKL}$ as well as a multiplication from the left by a compensating $SU(8)$ transformation depending on scalars as well as on 63 rigid parameters  $\Lambda_{J}{}^{L}$ and 70 rigid parameters $\Sigma_{IJKL}$
\begin{eqnarray}\label{Vtrans}
\mathcal{U}(y,\bar{y}; \Lambda, \Sigma) \cV(y')=\cV(y, \bar y) E^{-1}(\Lambda, \Sigma)\ ,
\end{eqnarray}
The compensating $SU(8)$ transformation with parameters depending on scalars and \E\, parameters in eq. \rf{Einverse} is
\be
{\cal U}(y(x), \bar y(x), \Lambda, \Sigma)=1+\frac{1}{2}\ P^{-1/2}\Delta(\Lambda,\Sigma)P^{-1/2}\equiv 1+\Upsilon\ , 
\label{U}\ee
where $\Delta(\Lambda,\Sigma)$ is given by
\begin{eqnarray}\label{Delta}
\Delta(\Lambda,\Sigma)=\{\Lambda,P\}+(y\bar{\Sigma}-\Sigma\bar{y})
-y(\bar{\Sigma}y-\bar{y}\Sigma)\bar{y}\ .
\end{eqnarray}
In \cite{Bossard:2010dq} in a different setup, it was also  found that the compensating $SU(8)$ transformation induced and 
by \E\, 
\be
g \equiv \exp {\Sigma} =  \left( \begin{array}{cc}  0  &  \Sigma_{ijkl}    \vspace{0mm}  \\
  \Sigma^{ijkl}  & 0 \end{array} \right) \; \; .
\label{La}\ee
has the following scalar-dependent form
\be
\delta^{SU(8)}\Big( \tanh(\Phi (x)/2) *{\Sigma} \Big)
\label{U1}\ee
where $\Phi(x) $ is defined in eq. \rf{gauge1} and the $*$ operation is associated with the   Baker-Campbell-Hausdorf formula for multiplication of two Lie algebra elements $ \exp \Phi(x)$ in  \rf{gauge1} and  $\exp {\Sigma}$ in \rf{La}.

Clearly,  one can see in both cases in \cite{Kallosh:2008ic,Bossard:2010dq} that the compensating $SU(8)$ transformation in \rf{U} and in \rf{U1}  depend on scalars fields $\phi_{ijkl}(x)$ and on \E\, parameters.

The reason that there is no superspace available corresponding to the unitary gauge with only physical scalars has to do with the fact that the on-shell superspace at the point where all $\theta$'s vanish represents the $x$-space supergravity equations of motion and their symmetries. These symmetries are either local or global; local ones, like Lorentz symmetry and internal symmetries,  become symmetries of the tangent space. The scalar field dependent $\cH$ symmetries are not compatible with the concept of the superspace, which has curved superspace general coordinate transformations and tangent space structure group.  Therefore, the unitary gauge in $\cN\geq 4$ supergravities cannot be represented in superspace.

The situation is different in $\cN=2$ case, where local $U(2)$ or a part of it can be removed from the structure group in superspace, see Appendix \ref{App:GIKOS}, where we present some known examples. Therefore $\cN=2 $ supergravity was considered in the harmonic or projective superspace, but not in case $\cN>2$ \cite{Galperin:2001seg,Kuzenko:2008ep,Kuzenko:2008ry}. An important difference between $\cN=2$ case and $\cN\geq 4$ supergravities is the presence of duality symmetry $\cG$ in $\cN\geq 4$  and absence of it in $\cN=2$.

In the on-shell harmonic superspace  \cite{Bossard:2011tq}, the additional coordinates $u^I{}_i$ were nevertheless added to $\cN\geq 4$ superspace, which, as we have shown, breaks local $\cH$ symmetry, and makes this superspace inconsistent.

\section{Harmonic Lagrangian and G-analyticity  break local  $\cH$ symmetry}

 The local $U(\csN)$ symmetry acts linearly on harmonic superspace superfields of a standard superspace \cite{Howe:1981gz} 
\be\label{deltachi}
\delta \bar \chi^{ijk} (x, \theta, \bar \theta)= \Lambda ^{[i}{}_l (x, \theta, \bar \theta) \bar \chi^{ljk]} (x, \theta, \bar \theta)\, , \ee
It is suggested in \cite{Bossard:2011tq}  that the new coordinates $u^I{}_i(t)$ are transformed by a local $U(\csN)$, which would mean that the parameters of transformations depend on local coordinates $(x, \theta, \bar \theta)$.  But in fact, this is inconsistent with the constraints on the harmonic superspace curvatures  \rf{Hcurv}, \rf{vanish} and the fact that $u^I{}_i$ depend only on local coordinates of the coset space. 

Therefore the harmonic superspace superfields like  
\be
\bar \chi^{Ijk} (x, \theta, \bar \theta, t)\equiv   u^I{}_i (t)\bar \chi^{ijk} (x, \theta, \bar \theta)\ee
are not covariant under local $U(\csN)$ symmetry.

The harmonic superspace curvatures in  \cite{Bossard:2011tq} are defined as a contraction of a  local $SU(\cN)$ tensor quadratic in spinorial superfields $\chi_{ijk \a}(x, \theta) $ and its conjugate with harmonic coordinates $u^1_i(t) , u^l_{\cN}(t)$.
These curvatures  are 
\be
(R^1_{ \a \, \dot \beta \cN})^1{}_1= (R^1_{ \a \, \dot \beta \cN})^{\cN}{}_{\cN} = {1\over 2} B_{\a \, \dot \beta}(x, \theta, \bar \theta , u)=
{1\over 2} u^1{}_l \, \Big (\bar\chi_{\dot\beta}^{lij}(x, \theta, \bar \theta) \chi_{\alpha\, k ij} (x, \theta, \bar \theta) \Big) \, u^k{}_{\cN} \ .
\label{Hcurv1}\ee
and the Lagrangian is
\be
{\cal L}= B_{\alpha \dot \beta}B^{\alpha \dot \beta} (x, \theta, \bar \theta, t)
\label{L}\ee
where
\be
 (B_{\alpha \dot \beta})^1{}_{\cN}\equiv  u^1_i u^l_{\cN} \bar \chi^{ijk}_{\dot \beta} \chi_{\alpha ljk}\, ,\quad (B^{\alpha \dot \beta})^1{}_{\cN}= \epsilon^{\alpha \beta} \epsilon ^{\dot \alpha \dot \beta} (B_{\beta \dot \alpha})^1{}_{\cN}
\label{sf}\ee
Under local $U(\cN)$ symmetry  \rf{deltachi} with $\cN^2$ independent local parameters $\Lambda ^i{}_l (x, \theta, \bar \theta)$,  the superfield $B_{\a \, \dot \beta}$  is not covariant
\be
\delta_{U(\csN)|_{\gf}} \, B_{\a \, \dot \beta}(x, \theta, \bar \theta, t)= \delta_{U(\csN)|_{\gf}}  \Big (u^1{}_l(t) \, \Big (\bar\chi_{\dot\beta}^{lij}(x, \theta, \bar \theta) \chi_{\alpha\, k ij} (x, \theta, \bar \theta) \Big) \, u^k{}_{\cN}(t)\Big )\neq 0
\ee
The reason for this is the fact that  the variation of $u^I{}_i(t)$ is not one which would cancel the variation of the superfield $\chi$. Namely, one would need a transformation
$
\delta u^I{}_i(t) =   u^I{}_k(t) \Lambda ^k{}_i (x, \theta, \bar \theta)
$
with $\cN^2$ independent parameters and linearly realized local $U(\cN)$ symmetry.  But this is not the case.
In particular, the actual transformation consistent with constraints on harmonic superspace curvatures requires that these symmetries in the coset direction are broken.

The Lagrangian is a contraction of a local $U(\cN)$ tensor quartic in superfields $\chi$, and the object depending on four $u^I_i(t)$'s
\be
{\cal L}= B_{\alpha \dot \beta}B^{\alpha \dot \beta}=T^{nl}{}_{im} \times {\cal L}^{im}{}_{nl}  (x, \theta, \bar \theta) \ .
\label{L1}\ee
The local $U(\cN)$ tensor is
\be
{\cal L}^{in}{}_{ml}  (x, \theta, \bar \theta) =\bar \chi^{ijk}_{\dot \beta} \chi_{mjk \a}\,  \bar \chi^{nop}_{\dot \alpha} \chi_{lop \beta}\,  \epsilon^{\alpha \beta} \, \epsilon ^{\dot \alpha  \dot \beta}\ee
and the object quartic in $u$'s is
\be
T^{nl}{}_{im}= u^1{}_i u^1{}_m u^{ n}{}_{\cN} u^{ l}{}_{\cN} \ .
\ee
It follows that the harmonic Lagrangian ${\cal L}$ breaks local  $U(\csN)$ symmetry in the coset directions of
${U(\cN)\over U(1)\times  U(\csN-2)\times U(1)}$ so that
\be
\delta_{U(\csN)|_{\gf}} {\cal L} =
\delta_{U(\csN)|_{\gf}} B_{\a \, \dot \beta}B^{\a \, \dot \beta} (x, \theta, \bar \theta, t)\neq 0
\ee
The superfield $B_{\alpha \dot \beta}$ and all its powers are Grassmann analytic according to \cite{Bossard:2011tq}. It means that
\be
D_\gamma ^1 B_{\alpha \dot \beta} = \bar D_{\dot \gamma \cN} B_{\alpha \dot \beta} =0 \ ,
\label{Gan1}\ee
where
\be
D_\gamma ^1 = u^1{}_i D_\gamma ^i\, \qquad  \bar D_{\dot \gamma \cN}=  D_{\dot \gamma i} u^i {}_{\cN} \ .
\ee
The proof of integrability condition for each $\cN\geq 4$ is given in Appendix A of \cite{Bossard:2011tq}  based on standard superspace geometry \cite{Howe:1981gz} with local $U(\cN)$ symmetry of the tangents space. Here $D_\gamma ^i$ and $D_{\dot \gamma i}$ transform under a local $U(\cN)$ symmetry according to positions of the indices $i$.

If we accept the claim in  \cite{Bossard:2011tq} that $u^1_i , u^l_{\cN}$ transform under local $U(\cN)$, we would think that the G-analyticity condition is invariant under local $U(\cN)$ symmetry. However, this is inconsistent with constraints on curvatures in the harmonic superspace. It means that the G-analyticity condition \rf{Gan1} breaks local $U(\cN)$ despite the integrability condition for the G-analyticity condition shown in eq. \rf{vanish} is satisfied.

Let us look at the G-analyticity of the Lagrangian \rf{L}. This Lagrangian is claimed to be duality-invariant G-analytic scalar superfield. The fact that it depends on spinorial superfield $\chi^{ijk}_{\a}(x, \theta) $ of the standard superspace, where this superfield is a superspace torsion, is indeed consistent with duality $\cG$ symmetry, under the condition that the local $U(\cN)$ structure group symmetry is valid. The G-analyticity of the Lagrangian means that
\be
D_\alpha ^1 {\cal L} = 
\bar D_{\dot \alpha \cN} {\cal L}=0  \ .
\ee
The analyticity condition of the Lagrangian can be presented in the form 
\be
D_\alpha ^1 {\cal L} = F^{nl}{}_{qim} \times (D_\alpha^q {\cal L}^{im}{}_{nl})  (x, \theta, \bar \theta) \ ,
\ee
where
\be
F^{nl}{}_{qim}= u^1{}_q u^1{}_i u^1{}_m u^{ n}{}_{\cN} u^{ l}{}_{\cN} \ .
\label{5u}\ee
It is also given by  a product of  local $U(\cN)$ tensor $(D_\alpha^q {\cal L}^{im}{}_{nl})  (x, \theta, \bar \theta)$ times a product of five $u$'s in eq. \rf{5u}. It is, therefore, not invariant under a local $U(\cN)$ symmetry.

If one would like to gauge-fix the local $U(\cN)$ symmetry in superspace, following procedure known in space-time, see Sec. \ref{gf}, without manifest supersymmetry,  one would have to build first a normal on-shell superspace breaking its tangent space local $U(\cN)$ symmetry. The compensating, superfield-dependent,  local $U(\cN)$ symmetry transformation would be required to preserve global $\cG$ symmetry. Even ignoring the fact that such an on-shell superspace with gauge-fixed local $U(\cN)$ symmetry was not constructed, we observe that at $\theta=0$ the compensating $U(\cN)$  transformation depends on scalars $\phi_{ijkl}(x)$. In superspace, it would depend on superfields, so the new coordinates $u_i^I(t)$ would be required to transform on $U(\cN)$ symmetry with superfield dependent parameters. Therefore, if a corresponding superspace with gauge-fixed local $U(\cN)$ symmetry would be constructed,  it would be inconsistent with harmonic coordinates $u_i^I(t)$: these cannot accommodate at $\theta=0$ spacetime field dependent $U(\cN)$ transformation. Thus, even in the unitary gauge, there is a clear inconsistency of the nonlinear harmonic superspace  \cite{Bossard:2011tq} for $\cN\geq 4$.

To summarize, the inconsistency with local $U(\cN)$ symmetry in  \cite{Bossard:2011tq} is that, on  one hand, this local symmetry is claimed to be not gauge-fixed. But actually, harmonic coordinates are treated by analogy with harmonic superspace coordinates \cite{Galperin:2001seg} or projective superspace coordinates in $\cN=2$ case. For example, in  \cite{Kuzenko:2008ep,Kuzenko:2008ry}, whose strategy is used in  \cite{Bossard:2011tq}, it is explained that the $u_i^+$  serve merely to totally symmetrize all $SU(2)$ indices $i$. Also, they present the explicit dependence of $u^+{}_i$ on two ${SU(2)\over U(1)}$ coset coordinates but only after the local $SU(2)$ symmetry was gauge-fixed in the coset directions. And since there was no $\cG$ duality symmetry, no compensating field dependent $SU(2)$ was required. It allowed a consistent construction of the harmonic or projective superspace in $\cN=2$ supergravity. This, however, is not possible in $\cN\geq 4$ supergravities.

It is not surprising therefore  that 12 graphs in $\cN=4$, $L=3$ \cite{Bern:2012cd} and 
82 diagrams in $\cN=5$, $L=4$ cancelled \cite{Bern:2014sna}: the $L=\cN-1$ harmonic superinvariants  in  \cite{Bossard:2011tq} break local $U(\cN)$ symmetry and, as a result of the algebra \rf{susy}, also a local nonlinear supersymmetry.

%\section{\boldmath Below critical superinvariants at $d>4$}

\section{Discussion}
In 4d  supergravity at $\cN\geq 4$   the critical loop order $L_{cr}={\cN}$  is the loop order where the geometric superinvariants \cite{Kallosh:1980fi,Howe:1980th} have both  local $\cH$ symmetry and global duality $\cG$ symmetry of the relevant on-shell superspace. Any subspace of the whole superspace with less fermionic directions and no additional coordinates, like harmonic ones,  breaks the local $\cH$ symmetry of the on-shell superspace.

In 4d $\cN\geq 4$ supergravity, an attempt was made in   \cite{Bossard:2011tq} to construct a harmonic on-shell superspace with the property that there are ${1\over \cN}$ BPS superinvariants built in a subspace of the whole superspace, where the number of fermionic directions instead of $4\cN$ is equal to $4(\cN-1)$.
This corresponds to candidate counterterms at $L=L_{cr} -1= \cN-1$.

In an on-shell harmonic superspace  \cite{Bossard:2011tq}, the integrability condition for the existence of Grassman analytic superfields, i.e., superfields independent on some of the fermionic directions, 
requires that harmonic superspace curvature components in a coset direction  ${H \backslash\cH}$ 
to vanish, namely
\be
(R_{\hat \a \, \hat \beta})^1{}_r= (R_{\hat \a \, \hat \beta})^r{}_{\cN}=(R_{\hat \a \, \hat \beta})^1{}_{\cN}=0
\label{vanishD}\ee
This is necessary for the G-analyticity condition 
\be
D_\gamma ^1 B_{\alpha \dot \beta} = \bar D_{\dot \gamma \cN} B_{\alpha \dot \beta} =0
\label{Gan}\ee
to be consistent with the harmonic superspace geometry.

However, harmonic superspace curvature components in  $H$-direction do not vanish 
\be
(R_{\hat \a \, \hat \beta})^1{}_1=R_{\hat \a \, \hat \beta})^{\cN}{}_{\cN}\neq 0
\label{NvanishC}\ee
since the harmonic superspace Lagrangian is built from these non-vanishing components of the curvature
\be
{\cal L }= (R_{ \a \, \dot \beta})^1{}_1 (R_{ \beta \, \dot \alpha})^1{}_1\epsilon^{\a\beta} \epsilon^{\dot \a \dot \beta} \, .
\ee
But to keep some curvatures vanishing and some not vanishing is only possible if local symmetries in the coset direction are broken: namely, there are  $4\cN-6$ constraints on  local $\cH=U(\csN)$ symmetry of the tangent space of the standard superspace \cite{Brink:1979nt, Howe:1981gz} with parameters 
$\Lambda^i{}_j (x, \theta, \bar \theta)$
\be
 u^1{}_i \Lambda^i{}_j (x, \theta, \bar \theta)u^j{}_r=u^r{}_i \Lambda^i{}_j (x, \theta, \bar \theta)u^j{}_{\cN}= u^I{}_1 \Lambda^i{}_j (x, \theta, \bar \theta)u^j{}_{\cN}=0
\label{ConLoc}\ee
And since local $U(\csN)$ symmetry of the tangent space is broken, it also means nonlinear local supersymmetry is broken, according to the algebra \rf{susy}.

At the linear level, in the unitary gauge with physical scalars only,  harmonic superspace $L= \cN-1$ superinvariants, constructed in  \cite{Bossard:2011tq}, are eligible superinvariants. At the linear level, the compensating $(S)U(\cN)$ transformation, required to preserve duality $\cG$ symmetry, is just a global field independent $\cH$ transformation. Therefore, harmonic coordinates are consistent with global $\cH$  symmetry. 
 But at the nonlinear level, the field-dependent compensating $SU(\cN)$ transformation is incompatible with harmonic coordinates, and the $L=L_{cr} -1= \cN-1$ superinvariants break nonlinear supersymmetry.

 The loop computation in $\cN=4$, $L=3$ \cite{Bern:2012cd}, and in $\cN=5$, $L=4$  \cite{Bern:2014sna} show that UV infinities supported by $\cN-1$ harmonic superinvariants cancel. We explained here that $\cN-1$ harmonic superinvariants break nonlinear $\cH$ symmetry and nonlinear supersymmetry. Therefore these symmetries explain  
  the absence of UV divergences in these cases.

\section*{Acknowledgement}
I am grateful to A. Linde,  H. Nicolai, R. Roiban,  and  Y. Yamada for useful discussions related to the current work and to E. Ivanov and E. Sokatchev for earlier discussions of harmonic superspace, in general. This work is 
 supported by SITP and by the US National Science Foundation grant PHY-2310429.
 
 \appendix 

\section{$\cN=2$ harmonic superspace and G-analyticity}\label{App:GIKOS}
Harmonic superspace constructions are known to be successful in super Yang-Mills theory with $\cN\leq 3$ \cite{Galperin:2001seg}. Minkowski space is the simplest example of a coset space ${G\over H}$ where $G$ is the Poincar\'e group and $H$ is the Lorentz group. The coset parameters are coordinates
$x^a$, $a=0,1,2,3$
\be
M^4={G\over H}= {\{L_{ab}, P_a\}\over \{L_{ab}\}}=x^a
\ee
The superspace is a space 
where  Poincar\'e supersymmetry algebra can be realized in the same way as the Poincar\'e algebra is realized in Minkowski space. Poincar\'e superalgebra has an outer automorphism group $U(\cN)$ (also known as R symmetry). The spinor generators $Q_\a^i$ and $\bar Q^{\dot \a i}$ transform according to the fundamental  representations ${\bf \cN} $ and ${\bf \bar \cN} $ of $U(\cN)$, respectively. 

The real superspace has coordinates
\be
\mathbb{R}^{4|4\cN} ={G\over H}= {\{L_{ab}, su(\cN), P_a, Q_\a^i, \bar Q^{\dot \a i} \}\over \{L_{ab}, su(\cN)\}}= (x^a, \theta_i^\a, \bar \theta^{\dot \a i} )
\ee
Here $H$ is the group of automorphisms $SL(2,C)\times SU(\cN)$. 

 For example, harmonic superspace for $\cN=2$ theories is obtained  by keeping only the $U (1)$ part of the automorphism group $SU (2)$ in the subgroup $H$ :
\be
\mathbb{HR}^{4+2|8} ={G\over H}= {\{L_{ab},  P_a, Q_\a^i, \bar Q^{\dot \a i},su(2) \}\over \{L_{ab}, u(1)\}}= (x^a, \theta_i^\a, \bar \theta^{\dot \a i},u^+_i)\equiv (X^A)\, , \qquad i=1,2
\ee
Two new even coordinates $ u^+_i$ correspond to the coset $SU(2)/U(1)$, a  two-dimensional sphere $S^2$ 
 and $u_i^-$ is its inverse, $u^{+i} u_i^-=1$. The spinor derivatives are defined as follows
\be
D_\a^{\pm} = u^{\pm}_i D_\a^i\, , \qquad \bar D_{\dot \a}^{\pm} = u^{\pm}_i \bar D_{\dot a}^i
\label{der1}\ee
The main advantage of the harmonic superspace is the existence of a new invariant subspace containing only half of the original Grassmann variables. The analytic superspace is defined as follows
\be
\mathbb{HA}^{4+2|4} ={G\over H}= {\{L_{ab},  P_a, Q_\a^i, \bar Q^{\dot \a i},T^{\pm \pm}, T^0 \}\over \{L_{ab}, T^0, Q_\a^+, \bar Q_{\dot \a}^+\}}
\ee
Here $T^{\pm \pm}, T^0 $ are generators of $SU(2)$ and 
\be
Q_\a^+=u^+_i Q_\a^i\, , \qquad \bar Q_{\dot \a}^+= u^+_i \bar Q_{\dot \a}^i\
\label{Q}\ee
The generators  $L_{ab}, T^0, Q_\a^+, \bar Q_{\dot \a}^+$ form a closed algebra. Derivatives in the harmonic directions are introduced
\be
\partial^{\pm\pm} = u^{\pm i} {\partial \over \partial u^{\mp i}}\, , \qquad \partial ^0 = u^{+ i} {\partial \over \partial u^{+ i}}- u^{- i} {\partial \over \partial u^{- i}}\ \
\ee
Under $SU(2)$ the new coordinates transform as follows
\be
(u^{\pm}_i)'= \Lambda _i{}^j (u^{\pm}_j)e^{\pm i \psi(\Lambda, u)}
\ee
where $\Lambda _i{}^j$ is an element of $SU(2)$ and $\psi$ is a local phase of induced $U(1)$ transformation, depending on $\Lambda$ and $u^+_i$. 

The condition of the Grassmann analyticity (G-analyticity) is now possible in the analytic harmonic superspace.  
\be
D_\a^+ \Phi(x_A, \theta^{\pm}_A, \bar \theta^{\pm}_A, u)=0\, , \qquad \bar D_{\dot\a}^+ \Phi(x_A, \theta^{\pm}_A, \bar \theta^{\pm}_A, u)=0 \ .
\ee
It means that there is a basis where the superfield  depends only on 1/2 of fermionic coordinates
\be
\Phi(x_A, \theta^{+}_A, \bar \theta^{+}_A, u^{\pm}_i) \ ,
\ee
where 
\be
x_A^a= x^a- 2i \theta^{(i} \sigma^a \bar \theta^{j)}u^+_i u^-_j \, , \qquad 
\theta^{\pm}_{A \, \a} = u^{\pm}_i \theta ^i_\a\, , \qquad \bar \theta^{\pm}_{A \, \dot \a} = u^{\pm}_i \bar \theta ^i_{\dot \a}\, .
\label{coord}\ee
Grassmann-analyticity defining derivatives $D_\a^+$ and $\bar D_{\dot\a}^+$ form a commutative algebra with the the harmonic derivative $D^{++}$.

Under $SU(2)$ transformations with parameters $\Lambda _i{}^j$ spinorial derivatives in eq. \rf{der}, supersymmetry charges in eq.  \rf{Q} and analytic space coordinates in eq. \rf{coord} are $SU(2)$ invariant.

In $\cN=2$ an example of the analytic superfield is an off-shell hypermultiplet. In a superspace without harmonic variables it is $q^i (X)$. on-shell $D_\a^{(i }q^{j)} (X)=0$, but off-shell there is a harmonic superfield which is G-analytic
\be
q^+ (X, u) = q^i(X) u_i^+\, , \qquad D_\a^+ q^+(X, u) = \bar D_{\dot \a}^+ q^+(X, u)=0 \ .
\label{hyper} \ee
Note that the superfield $q^+ (X, u)$ and its G-analyticity conditions defined in eq. \rf{hyper}  are invariant under  $SU(2)$ transformations with superspace-independent parameters $\Lambda _i{}^j$.

%\section{$\cN=2$ supergravity superspace    }\label{App:N2}
%\subsection{On-shell and off-shell superspace in $\cN=2$ Poincar\'e supergravity} \label{App:N2d}
Conformal supergravities are associated with the superconformal algebra $SU(2, 2 | \cN)$; at the nonlinear level, they are available only at $\cN\leq 4$. In superspace, conformal supergravities  were described in \cite{Howe:1981gz} for 4d
$\cN\leq 4$ by explicitly gauging $SL(2, \mathbb{C}) \times U(\cN )$ and identifying the relevant constraints on the torsion of curved superspace. It is therefore not accidental that, for example,  conformal $\cN=2$ supergravity has a local in superspace $U(2)$ symmetry, which is in the structure group of the conformal superspace.

In case of $\cN=2$ Poincar\'e supergravity is obtained from conformal one in \cite{Howe:1981gz}, by de-gauging $U(2)$ symmetry, so that the tangent space 
is just $SL(2, \mathbb{C})$. 
The following is a short version of a derivation of Poincar\'e $\cN=2$ supergravity superspace in \cite{Howe:1981gz}. One begins with $SL(2, \mathbb{C}) \times U(2 )$ connections in conformal supergravity 
\be
\hat \Omega_A{}^B = \Omega_A{}^B +\Sigma_A{}^B \ ,
\ee
where $\hat \Omega_A{}^B$ is $SL(2, \mathbb{C}) \times U(2 )$ connection and $\Omega_A{}^B$ is $SL(2, \mathbb{C})$ connection and  proceed with de-gauging by requiring  restrictions on $\Sigma_A{}^B$ so that the superspace has no internal symmetry anymore. First, the local $U(1)$ is gauge-fixed, and afterwards also the local $SU(2)$. One of the important restrictions explaining the difference between off-shell and on-shell superspace is 
\be
\Sigma^{ij}_{\alpha \, k} = 2 \delta^i_k \rho^j_\alpha - \delta^j_k \rho^i_\alpha \ .
\ee
When constraints on torsion were imposed, as well as de-gauging constraints, the superspace Bianchi identities were solved consistently. In the final off-shell superspace with the tangent group $SL(2, \mathbb{C})$ there are two covariant superfields
\be
M_{\alpha \beta} (x, \theta, \bar \theta) \qquad {\rm and} \qquad \rho^i_\alpha (x, \theta, \bar \theta)
\ee
A superfield equation of motion is
\be
\rho^i_\alpha (x, \theta, \bar \theta)=0
\ee
Thus, all non-vanishing components of torsion and curvatures on-shell depend on one superfield $M_{\alpha \beta}$. There is no obstruction to adding harmonic coordinates to this superspace since there is no local $SU(2)$ symmetry, and all geometric Bianchi identities were solved consistently. This will be different in the case of the on-shell  $\cN\geq 4$ supergravities, where the local $(S)U(\cN)$ has a different nature and global duality symmetries $\cG$ play an important role.

%\subsection{Harmonic and Projective $\cN=2$ superspace}\label{App:B2}

Harmonic  $\cN=2$ superspace for  Poincar\'e supergravity is described in \cite{Galperin:2001seg}, starting with a  conformal $\cN=2$ superspace. To gauge-fix extra local symmetries, the compensator superfields are introduced. They are used to gauge fix the local $SU(2)$ so that the gauge-fixed local $SU(2)$ is compatible with harmonic coordinates $u_i^{\pm}$, i=1,2 in $\cN=2$ Poincar\'e supergravity.  

The construction of  $\cN\geq 4$ harmonic superspace in  \cite{Bossard:2011tq}, where $L=\cN-1$ counterterms are claimed to exist, follows the strategy of $\cN=2$ projective superspace 
in \cite{Kuzenko:2008ep,Kuzenko:2008ry}.
We will explain the reason why the construction in projective superspace in \cite{Kuzenko:2008ep} and the consequent studies of the action of  $\cN=2$ supergravity in \cite{Kuzenko:2008ry} are consistent with the presence of isotwistor variables $u^+_i$,  $i=1,2$. These isotwistor variables $u_i^+$ are considered to be inert with respect to supergravity gauge group $SU(2)$. It is stressed that $u_i^+$ are constant.

In \cite{Kuzenko:2008ep,Kuzenko:2008ry} the fermionic derivatives are defined as 
\be
D^+_\alpha := u_i^+ D_\alpha ^i, \, D^-_\alpha := u_i^- D_\alpha ^i
\label{contr}\ee
But since $u^{\pm}_i$ are constant, they cannot transform under local $SU(2)$.
The reason why this is not a problem in $\cN=2$ supergravity in \cite{Kuzenko:2008ep,Kuzenko:2008ry} is the following. The action derived in \cite{Kuzenko:2008ep} is invariant under the super-Weyl transformations generated by a covariantly chiral parameter $\sigma$ where $\bar D_{\dot \a}^i \sigma =0$. It also has a local $U(2)$ symmetry. To get $\cN=2$ Poincar\'e supergravity, one must gauge-fix various extra local symmetries, including the local $SU(2)$. 

The action of conformal supergravity in eq. (1.2) of \cite{Kuzenko:2008ry}  depends on a superconformal compensator, an off-shell tensor multiplet described by a symmetric real superfield $H^{ij} (x, \theta)$. When $H^{ij}$ is constrained \be
H^{ij}= -{i\over 2} (\sigma_1)^{ij}\, G\, ,  \qquad (\sigma_1)^{ij}=\left(\begin{array}{cc}0 & 1 \\1 & 0\end{array}\right)^{ij} 
\ee
the action does not have local $SU(2)$ symmetry anymore, only the local U(1) is left. The local U(1) gauge connection is $\Phi_\alpha^{i\, jk}= i \,\Sigma^i_\alpha (\sigma_1)^{ij}$. One can further use the de-gauging procedure for $\cN=2$ as in \cite{Howe:1981gz}, where one makes a choice 
$
\Sigma^i_\alpha =0
$ so that $SU(2)$ gauge symmetry is fixed and only global $SU(2)$ remains. This is consistent with the fact that the isotwistor coordinates $u_i^+$ and their inverse are inert under local $SU(2)$.

Therefore in $\cN=2$ supergravity a contraction of the type \rf{contr}
is acceptable, local $SU(2)$ is gauge-fixed in superspace.  
\section{Superinvariants below critical loop order in $d>4$}\label{App:C}
%\subsection{Critical loop order}
Dimensional analysis in superspace to establish the value of $L_{cr}$ starting from which the geometric superinvariants exist was performed in   \cite{Kallosh:2023css}. Here we reproduce the result for $L_{cr}$  obtained in  \cite{Kallosh:2023css} in a slightly more convenient form.
A dimension of the  gravitational coupling is 
$
{\rm dim} \Big [{1\over \kappa^2} \Big]= {\rm dim}  [M_{Pl}^2] = d-2  
$.
For integer dimensions $d\geq 4$ we determine 
 the critical loop order defined by an on-shell counterterm which is a geometric whole superspace integral with the Lagrangian invariant under global $\cG$ and local $\cH$.
\be
S_{cr} = \kappa^{2 \, (L_{cr}-1)} \int d^{4\cN} d^d x \, det\,  E\, \cL(x, \theta) \ .
\ee  
In such case, the minimal dimension superfield $\cH$-invariant Lagrangian depends on 4 spinor dimension 1/2 superfields and their superspace derivatives. 
\be
{\rm dim} \Big [\cL(x, \theta) \Big]= 2+n  \ , \qquad n > 0
\ee
In the presence of $n$, one can always make $ L_{cr}$ an integer.
The reason $n$ is positive is that all covariant superspace derivatives, as well as superfield-dependent torsions and curvatures,  have positive dimension. The counting goes as follows
 \be
{\rm dim} [\kappa^2] \, (L_{cr}-1) + {\rm dim}\,  [\rm full \, \, superspace] +{\rm dim}\,[\cL]=0 \, , 
\ee
so that the action is dimensionless, and we find
\be
-(d-2) (L_{cr}-1)+ 2\cN -d+2+n =0 \ ,
\ee
It follows that
\be
 L_{cr}= {2\cN +n \over (d-2)} \ , \qquad n\geq 0
\ee
In each case, we need to find a minimal value of $n$, which makes $ L_{cr}$ an integer. The result for the case of maximal supergravity is 
\bea 
 &d=4, \, L_{cr}=8 :  \quad &\kappa^{14} \int d^4\, x \, D^{10} R^4 +\dots \,  \quad  \quad \quad n=0 \cr
&d=5,  \, L_{cr}=6 :  \quad &\kappa^{10} \int d^5\, x \, D^{12} R^4+\dots \,    \quad \quad    \quad n=2 \cr
& d=6,  \, L_{cr}=4 :  \quad &\kappa^{6} \int d^6\, x \, D^{10} R^4+\dots \,    \, \, \quad \quad  \quad n=0\cr
& d=7,  \, L_{cr}=4 :  \quad &\kappa^{6} \int d^7\, x \, D^{14} R^4+\dots \,    \, \quad \quad   \quad n=4 \cr
& d=8,  \, L_{cr}=3 :  \quad &\kappa^{4} \int d^8\, x \, D^{12} R^4+\dots \,    \, \quad \quad   \quad n=3\cr
& d=9,  \, L_{cr}=3 :  \quad &\kappa^{4} \int d^9\, x \, D^{15} R^4+\dots \,    \, \quad \quad   \quad n=5  \nonumber
\eea
In \cite{Bern:1998ug,Bern:2007hh,Bern:2008pv,Bern:2009kd,Bern:2012uf,Bern:2018jmv} 
the first divergences in maximal supergravities have been found at $5\leq d\leq9$
\bea
&&d = 9, \qquad L_{UV} = 2 <  L_{cr}=3 \cr
&&d = 8, \qquad L_{UV} = 1 <  L_{cr}=3 \cr
&&d = 7, \qquad L_{UV}= 2 < L_{cr}=4\cr
&&d=6,   \qquad L_{UV}=3 <L_{cr}=4\cr
&&d=5,   \qquad L_{UV}=5 < L_{cr}=6
\label{compare}\eea
It appears that in all cases where we have the information about UV divergences in $d\geq 5$ from loop computations, the UV divergences were discovered at the loop order below what we have defined as $L_{cr}$.  According to our assertion, the nonlinear supersymmetry invariants cannot
be constructed unless they come from a whole superspace integral. Therefore, it follows from \rf{compare} that the UV divergences and the relevant CTs in all these maximal supergravities above 4d break full nonlinear supersymmetry and local $\cH$-symmetry.
%\subsection{Harmonic superinvariants below critical order }
At the linearized level 
in $d>4$, the relevant ${\cG\over \cH}$ supergravity superinvariants  below critical loop order were constructed in harmonic superspace in \cite{Bossard:2009sy,Bossard:2014lra,Bossard:2014aea,Bossard:2015uga}. In all cases, they represent the linearized version of the superinvariants at loop level below 
$
 L_{cr}={2\cN +n \over (d-2)}, \, n>0
$
The analysis of harmonic superinvariants below critical order performed in this paper for $d=4$ can now be applied to $d>4$. We focus on maximal supergravities in each dimension. In all cases, the physical scalars are in the coset space
\be
{\cG\over \cH} \quad : \quad  {E_{6(6)}\over USp(8)} \, , \quad {E_{5(5)}\over USp(4)\times USp(4)}\, , \quad {E_{4(4)}\over USp(4)}
\, , \quad {E_{3(3)}\over U(2)} \, , \quad {GL(2)\over SO(2)}
\ee 
 for $d=5,6,7,8,9$ respectively. There are Lagrangians with  local $\cH$ symmetry:  
\be
USp(8)\, , \quad  USp(4)\times USp(4)\, , \quad  USp(4)\, , \quad
U(2)\, , \quad SO(2)
\ee
 for $d=5,6,7,8,9$ respectively. Harmonic variables in all cases parametrize the coset space ${\cH\over H}$ where $H$ is a subgroup of $\cH$.
 
 The geometric whole superspace invariants at $ L_{cr}= {2\cN +n \over (d-2)}$ have volume of integration over all fermionic directions with local $\cH$-symmetry. Harmonic coordinates allow a smaller subspace of fermion variables by imposing Grassmann analyticity conditions on superfields.
 
 For example, in maximal $\cN = (2, 2)$ supergravity in six dimensions the relevant harmonic variables parametrize one of $USp(4)\times USp(4)$ as follows \cite{Bossard:2014lra}:
$
 {USp(4)\over U(2)}
$.
 This provides the Grassmann analytic superfields and linearized superinvariant describing  3-loop UV divergence of d=6 maximal supergravity. As in 4d case, the fact that harmonic variables break local $USp(4)\times USp(4)$ symmetry can be established using the superspace geometry. This means that there is no nonlinear generalization of the ${1\over \cN}$ BPS superinvariant presented in eq. (3.32) in \cite{Bossard:2009sy}. In fact, there is no claim that such a harmonic below critical order superinvariant has a nonlinear generalization in any of the papers in \cite{Bossard:2009sy,Bossard:2014lra,Bossard:2014aea,Bossard:2015uga}. Moreover, concerning 6d and 8d maximal supergravities, it is explained that there is an obstruction for the generalization of some of harmonic invariants to a nonlinear level. One can also check, case by case, that all harmonic superinvariants below critical order have no nonlinear generalization.

The reason why in 4d there was a claim in  \cite{Bossard:2011tq} that superinvariants below critical order exist in harmonic superspace compatible with nonlinear supersymmetry is the fact that harmonic Lagrangian depended on geometric spinorial superfields and had dimension 2. But in all $d>4$ cases,  harmonic Lagrangians for superinvariants below critical order in \cite{Bossard:2009sy,Bossard:2014lra,Bossard:2014aea,Bossard:2015uga} have dimension zero, they are not geometric and clearly exist only at the linear level.

To summarize,  harmonic superinvariants at $L<L_{cr}$ are useful to describe the UV divergences that did occur at $d>4$. None of these harmonic superinvariants with Grassmann-analytic superfields were claimed to exist at the nonlinear level.  In all cases, they are at $L<L_{cr}$ and show that local $\cH$ symmetry and nonlinear supersymmetry are broken by quantum corrections. This means that computations in \cite{Bern:1998ug,Bern:2007hh,Bern:2008pv,Bern:2009kd,Bern:2012uf,Bern:2018jmv}  have already established that these local symmetries are anomalous, and therefore, all $d>4$ maximal supergravities do not have consistent perturbative quantum field theory corrections.

Only in 4d case it is not known yet if  perturbative quantum field theory corrections are consistent with all symmetries of the theory.  The 4-loop cancellation of UV divergence in $\cN=5$ indicates a possibility that 4d maximal supergravity might be UV finite perturbatively. But more loop computations will be necessary to find out the actual UV properties of higher loop corrections in 4d maximal supergravity.

\bibliographystyle{JHEP}
\bibliography{refs}

\end{document}